# Losses of Interface Waves in Plasmonic and Gyrotropic Structures

A. Schuchinsky

University of Liverpool, L3 5TQ, Liverpool, UK, a.schuchinsky@liverpool.ac.uk

*Abstract* – The loss mechanisms of slow interface waves in the layered resonant media are examined and illustrated by the examples of (i) surface plasmon polaritons in an isotropic plasma layer, (ii) magnetoplasmons in magnetised plasma and (iii) spin waves in ferrimagnetic layers. It is shown that losses of all these interface waves grow at the same rate of $\text{Im}\gamma \sim \text{Re}\gamma^3$, where $\gamma$ is the wavenumber. These abnormal losses are caused by vortices of the power flow of the interface waves near their resonance cut-off. The basic properties of the slow interface waves discussed in the paper are inherent to the waves of hyperbolic type in the layered resonant media.

## I. INTRODUCTION

Slow electromagnetic waves guided by the layers and interfaces of the resonant plasmonic and ferrimagnetic structures represent a distinct class of the surface waves. These waves are of *hyperbolic* type and exist only in the finite frequency bands, being resonantly absorbed at their upper frequency cut-offs. Their main properties are discussed in this paper by the examples of interface waves (IWs) such as surface plasmon polaritons [1]-[7], magnetoplasmons [8]-[13] and spin waves [14]-[17]. The slow bulk waves (BWs) like magnetoplasmons and spin waves exist when the magnetic bias has components directed along the wave propagation or normal to the guiding interface. The mechanisms of the BW propagation, dissipation and power flow are somewhat similar to the IWs and they are not discussed in detail here.

The properties, functionality and applications of the slow IWs and BWs in the resonance media have been extensively studied in the literature, see e.g., [1]-[17] and references therein. In contrast to the conventional surface waves guided by dielectric layers, IWs and BWs are slower than the plane waves in the constituent media of the guiding structure. Therefore, these waves of the hyperbolic type cannot be described by a basic superposition of plane waves. The waves in the hyperbolic metamaterials have recently attracted increased attention [18]-[26]. Their various applications have been proposed and explored in the literature [19], [20], [23]-[26], including the promise of enhancing the sub-wavelength resolution [19], [20] and realising "slow light" [27], [28]. However, the published practical demonstrators exhibited high losses, which were notably higher than in the conventional devices based on dielectric waveguides and resonators, and optical fibres. Therefore, the detailed analysis of the loss mechanisms of the slow IWs and BWs in *imperfect* hyperbolic media is essential for their practical use.

The effects of the medium losses on the properties of IWs and their Poynting vector are examined and quantified in this work. The properties of the IW propagation and dissipation are elucidated with the examples of the waves guided by the interfaces of dielectric layers with isotropic and magnetised plasma and ferrimagnetic layers. The paper scope includes the analysis of
  - the basic modes of the slow IWs,
  - the dispersion and attenuation characteristics of the IWs, and
  - the effect of the power flow vorticity on the IW propagation and resonance losses.

The main properties of the IWs are discussed by the three examples:
  (i) Surface plasmon polaritons (SPPs) in isotropic plasma layer,
  (ii) Magnetoplasmons (MPs) in tangentially magnetised plasma layer and
  (iii) Spin waves (SWs) in tangentially magnetised ferrimagnetic layer.

The results of this work demonstrate that the anomalous losses are inherent to the IWs. In contrast to the conventional surface waves, the IWs are of the hyperbolic type, and their attenuation constants are proportional to a *cube* of their propagation constants. This is why the IWs exhibit strong attenuation in the proximities of their high frequency resonance cut-offs. It is shown that both the slow propagation and the high losses of the IWs are intrinsically linked to vorticity of the power flow of the IWs in the hyperbolic medium, and the examples of the three types of IWs illustrate this effect.

The paper is organised as follows. The canonical 3-layer structure, used for the analysis of the SPPs, MPs and SWs, is described in Section II. SPPs in thin metallic layers are discussed in Section III. MPs in the tangentially magnetised plasma layer are considered in Section IV, and spin waves in Section V. The main properties of the IWs, their dissipation and power flow are summarised in Conclusion.

## II. CANONICAL STRUCTURE

The basic 3-layer planar structure shown in Fig. 1 is used for the study of IWs bound to the central layer of thickness $a_0 = 2a_c$. Two isotropic dielectric layers have relative permittivities $\varepsilon_{1,2}$ and thicknesses $a_{1,2}$. The whole structure is bounded by the perfect electric conductor[1] (PEC) walls located at $y = (a_1+a_c)$, $-(a_2 + a_c)$. The central layers are of the following types

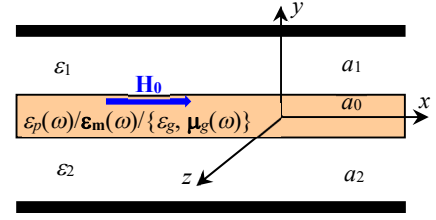

Fig. 1. Canonical 3-layer structure in PEC enclosure. Magnetic bias **H₀** is applied only to plasma and ferrimagnetic centre layers.

(i) an isotropic plasmonic layer with Drude scalar permittivity $\varepsilon_p(\omega)$,
(ii) a gyrotropic plasma slab, magnetised along the $x$-axis and described by Voigt permittivity tensor $\boldsymbol{\varepsilon_m}(\omega)$,
(iii) a ferrimagnetic layer, magnetised along the $x$-axis and described by a scalar permittivity $\varepsilon_g$ and Polder permeability tensor $\boldsymbol{\mu}_g(\omega)$.

These types of the central layers have negative effective permittivity or permeability in the finite frequency bands limited by the intrinsic resonances of the medium. They support propagation of the IWs of hyperbolic type, which have the high frequency resonance cut-offs. In the proximities of the cut-off frequencies, losses of the IWs rapidly grow in the resonant manner, and the attenuation constants of the IWs are proportional to a *cube* of the propagation constants. It is shown below that the high losses of the IWs is their inherent property related to vorticity of their Poynting vector. The main mechanisms of losses and power flow of the IWs are elucidated by the examples of SPPs, MPs and SWs in isotropic and gyrotropic (magnetised plasma and ferrimagnetic) layers.

## III. SURFACE PLASMON POLARITONS IN PLASMONIC LAYER

Let us consider the 3-layer structure shown in Fig. 1 where the central layer is the isotropic plasma with Drude permittivity $\varepsilon_p(\omega)$ defined as

$$\varepsilon_p(\omega) = \varepsilon_L \left[ 1 - \frac{\omega_p^2}{\omega \Omega} \right] \quad (1)$$

where $\Omega = \omega - j\nu$, $\omega$ is angular frequency, $\omega_p$ and $\nu$ are the plasma and collision frequencies, respectively, and $\varepsilon_L$ is the background permittivity.

Eigenwaves in a planar structure shown in Fig. 1 include TE and TM waves [29]. TE waves with the field components $E_x$, $H_y$ and $H_z$ are the ordinary surface waves. They are the guided modes of the plasmonic layer only at $\mathrm{Re}\,\varepsilon_p(\omega) > \max(\varepsilon_1, \varepsilon_2)$ and layer thickness $a_0$ about a half wavelength. For thin layers, these conditions are fulfilled only at very high frequencies $\omega \gg \omega_p$. Therefore, TE waves are not considered here.

TM waves with the field components $H_x$, $E_y$ and $E_z$ are the extraordinary modes of the plasmonic layer. They have been extensively studied in the literature, see, e.g., [2]-[7], [10] and references therein. The dispersion equation (DE) of TM waves with the wave propagator $\exp\{j(\omega t - \gamma z)\}$ is readily obtained by enforcing the boundary conditions of the tangential field continuity at the layer interfaces and at PEC enclosure. The DE can be presented as follows

$$K_1(\gamma) K_2(\gamma) - \left[ \frac{\beta_p}{\sinh(\beta_p a_0)} \right]^2 = 0 \quad (2)$$

where $K_m(\gamma) = \varepsilon_p V_m + \beta_p \coth(\beta_p a_0)$, $V_m = \frac{\beta_m}{\varepsilon_m} \tanh(\beta_m a_m)$, $\beta_m = \sqrt{\gamma^2 - k_0^2 \varepsilon_m}$, $m = 1, 2$; $\beta_p = \sqrt{\gamma^2 - k_0^2 \varepsilon_p(\omega)}$;

$\gamma$ and $k_0$ are longitudinal and free space wavenumbers, respectively. The main features of the TM wave dispersion and attenuation are illustrated in Fig. 2, obtained by numerical solution of (2).

Spectrum of the fundamental TM modes in the plasmonic layer includes the conventional surface waves and SPPs. The surface waves are the bound modes only at $\omega > \omega_p$, see Fig. 2, when the plasmonic layer acts as a dispersive dielectric waveguide with $\mathrm{Re}\,\varepsilon_p(\omega) > \max(\varepsilon_1, \varepsilon_2)$. The propagation constants of these surface waves vary in the range $\sqrt{\max(\varepsilon_1, \varepsilon_2)} < \mathrm{Re}\,\gamma/k_0 < \mathrm{Re}\,\sqrt{\varepsilon_p(\omega)} < \sqrt{\varepsilon_L}$. At $0 < \mathrm{Re}\,\varepsilon_p(\omega) < \max(\varepsilon_1, \varepsilon_2)$, the eigenwaves are not bound to the plasmonic layer and leak into a dielectric layer with a higher permittivity.

At frequencies $\omega < \omega_p$, $\mathrm{Re}\,\varepsilon_p(\omega) < 0$ and only SPPs with wavenumbers $\mathrm{Re}\,\gamma_m > k_0 \sqrt{\varepsilon_m}$, $m = 1, 2$ are guided by the plasmonic layer. SPPs are the IWs and their fields decay exponentially from the layer surfaces. In the case of dielectric layers with unequal permittivities, $\varepsilon_1 \neq \varepsilon_2$, the dispersion characteristics of SPPs and cut-off frequencies $\omega_{r1}$ and $\omega_{r2}$ differ. From the definition of $\beta_p$ and $\beta_m$ in (2), it is evident that magnitude of the SPP fields decay much faster in the plasmonic layer than in the surrounding it dielectric layers. Therefore, the effect of the non-guiding interface is exponentially small and the last term in (2) can be neglected. Then the DEs of the two SPPs at the opposite interfaces are separated

$$\frac{K_m(\gamma)}{\varepsilon_p} = \frac{\beta_m}{\varepsilon_m}\tanh(\beta_m a_m) + \frac{\beta_p}{\varepsilon_p}\coth(\beta_p a_0) \simeq 0, \quad m = 1, 2. \quad (3)$$

where $\beta_m$ and $\beta_p$ are defined in (2). The dispersion characteristics of SPPs in Fig. 2 show that away from the SPP resonances at $\omega_{r1}$ and $\omega_{r2}$, the dispersion curves of $SPP_1$ and $SPP_2$ are fairly well correlated with those of the $SPP'_1$ and $SPP'_2$ in the lossless plasmonic layer. However, this is not the case at $\omega_{r2} < \omega < \omega_p$, as the backward $SPP'_3$ of the lossless plasmonic layer, becomes a strongly attenuated complex wave $SPP_3$ in the lossy plasmonic layer. Thus, losses qualitatively alter the properties of SPPs in the proximities of the plasmonic resonances $\omega_{r1,2}$.

---
[1] The PEC enclosure is used here for a sole purpose of making the eigenwave spectrum discreet. This facilitates a rigorous analysis of the complete spectrum including complex and leaky waves unbounded to the centre layer.

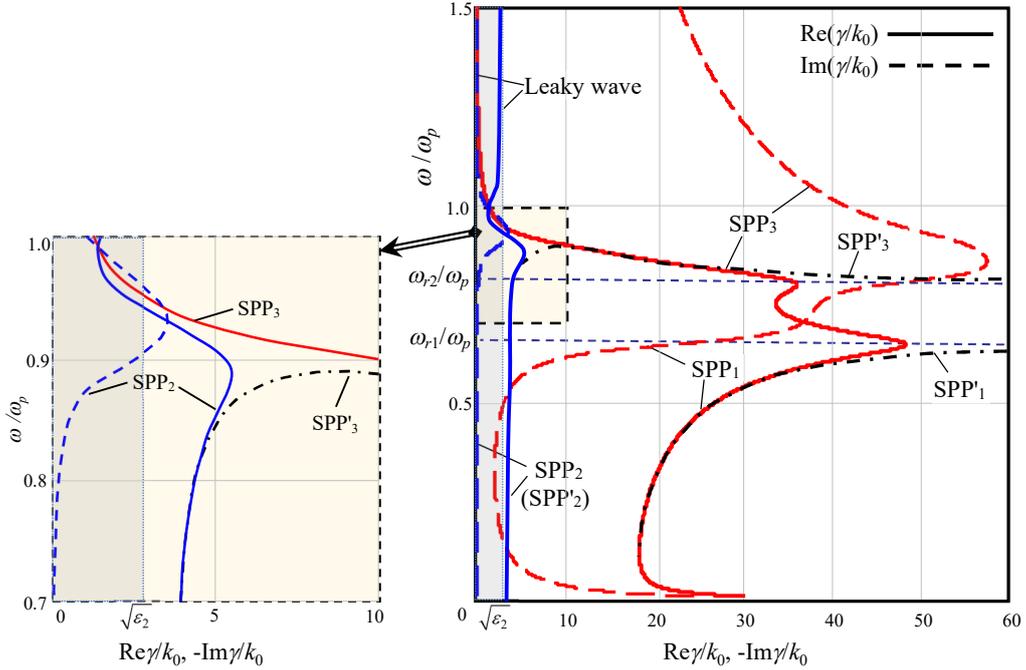

Fig. 2. Dispersion (solid lines) and attenuation (dashed lines) of SPPs guided by isotropic metallic film in a planar waveguide of Fig. 1. The structure parameters: $a_1 = a_2 = 40$ μm, $a_0 = 20$ μm, $\varepsilon_L = 13.1$, $\varepsilon_1 = 18.0$, $\varepsilon_2 = 7$, $\omega_p = 135$ GHz, $v = 0.05\omega_p$. Dotted and dash-dotted lines show the dispersion characteristics of $SPP'_1$ and $SPP'_3$ in the lossless case ($v = 0$). Side panel shows the magnified characteristics of the SPPs near the cut-off frequency $\omega_{r2}$. The dispersion curves in the grey-shaded area at $\mathrm{Re}\,\gamma/k_0 < \sqrt{\varepsilon_2}$ are of the leaky waves, which are not bound to the plasmonic layer.

To quantify the effect of losses, let us examine the asymptotic solutions of (3) at $\mathrm{Re}\,\gamma_m \gg k_0\sqrt{\varepsilon_m}$, $m = 1, 2$. The closed form approximations of the SPP propagation, $\mathrm{Re}\,\gamma_m$, and attenuation, $\mathrm{Im}\,\gamma_m$, constants at each interface are

$$\mathrm{Re}\,\gamma_m = k_0\sqrt{\varepsilon_m \left|\frac{\varepsilon_p}{\varepsilon_p + \varepsilon_m}\right|\frac{1+\chi_m}{2}} + O\left(\frac{k_0^2}{|\gamma_m|^2}\right)$$

$$\mathrm{Im}\,\gamma_m = |\mathrm{Re}\,\gamma_m|^3 \frac{2\,\mathrm{Im}(\varepsilon_p)}{k_0^2 |\varepsilon_p|^2 (1+\chi_m)} + O\left(\frac{k_0^2}{|\gamma_m|^2}\right) \qquad m = 1,2 \quad (4)$$

where $\chi_m = \dfrac{|\varepsilon_p|^2 + \varepsilon_m \mathrm{Re}\,\varepsilon_p}{|\varepsilon_p||\varepsilon_p + \varepsilon_m|}$, $m = 1, 2$. In the proximity of plasmonic resonances, $\chi_m = \dfrac{v}{\omega_p}\left(1 + \dfrac{\varepsilon_L}{\varepsilon_m}\right)\left[1 + O\left(\dfrac{v^2}{\omega_p^2}\right)\right]$,

i.e., $\chi_m$ is small at $v(1+\varepsilon_L/\varepsilon_m) < \omega_p$. Equations (4) reveal the *fundamental property* of SPPs that their attenuation has the anomalous growth rate of $\mathrm{Im}\,\gamma_m \sim (\mathrm{Re}\,\gamma_m)^3$ near the resonance cut-off, and this clearly seen in Fig. 2. This is why the SPPs decay much faster than the conventional surface waves with the attenuation constants $\sim \mathrm{Re}\,\gamma_m$.

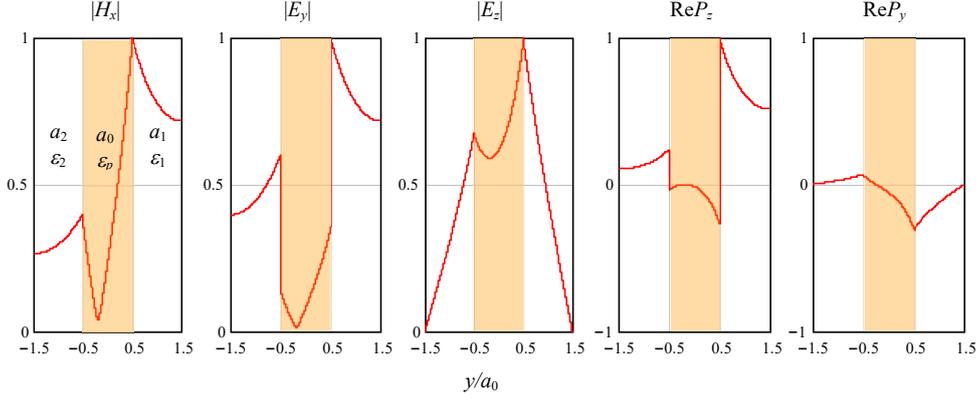

Fig. 3. Normalized cross-sectional distributions of the SPP fields $H_x$, $E_y$, $E_z$, and Poynting vector $P_z$, $P_y$ components in gold film located at $|y|<a_0/2$ (shaded area) at frequency $f = 591$ THz; $\gamma = k_0\,(1.87\text{-}j0.38)$. Layer thicknesses: $a_0 = a_1 = a_2 = 50$ nm and permittivities: $\varepsilon_1 = 1.5$ and $\varepsilon_L = \varepsilon_2 = 1$.

The abnormal rate of the SPP dissipation is directly related to vorticity of the power flow. Fig. 3 shows cross-sectional distributions of the fields and Poynting vector in the asymmetric structure containing a plasmonic layer (gold film with $\mathrm{Re}\,\varepsilon_p(\omega) < 0$) between lossless dielectric layers with permittivities $\varepsilon_1=1.5$ and $\varepsilon_2=1$. The normalised longitudinal component of Poynting vector, $P_z$, in dielectric and plasmonic layers has the following form

$$P_z(y) = \frac{\eta_0}{k_0} Q^2\, \mathrm{Re}\begin{cases} \dfrac{\gamma}{\varepsilon_m}\left[\left(1-(-1)^m W_e(\varepsilon_p,0)\right)\dfrac{\cosh\beta_m\left(a_m+a_c+(-1)^m y\right)}{\cosh\beta_m a_m}\right]^2, & y(-1)^{m-1} > a_c,\ m=1,2 \\[1em] \dfrac{\gamma}{\varepsilon_p(\omega)}\left[\dfrac{\cosh\beta_p y}{\cosh\beta_p a_c} + W_e(\varepsilon_p,0)\dfrac{\sinh\beta_p y}{\sinh\beta_p a_c}\right]^2, & |y|\le a_c \end{cases} \qquad (5)$$

where $\eta_0$ and $k_0$ are the free space impedance and wavenumber, respectively, $Q$ is the normalised magnitude of the magnetic field, and

$$W_e(\upsilon,\varpi) = \frac{\upsilon\left[V_2(\gamma)-V_1(\gamma)\right]-2\gamma\varpi}{\upsilon\left[V_2(\gamma)+V_1(\gamma)\right]+2\beta_p \coth(\beta_p a_c)} \qquad (6)$$

The distinctive feature of the Poynting vector distribution is that at $\mathrm{Re}\,\varepsilon_p(\omega) < 0$, its longitudinal component $P_z(y)$ inside and outside the plasmonic layer has opposite signs, see (5). This is particularly evident in the case of $V_1(\gamma) = V_2(\gamma)$ when $W(\varepsilon_p, 0) = 0$. When permittivities of the dielectric layers are unequal, $\varepsilon_1 \ne \varepsilon_2$, the field and power flow distributions of the SPPs guided by the opposite interfaces of the plasmonic layer are asymmetric because $V_1(\gamma) \ne V_2(\gamma)$ in (6). Plots of $\mathrm{Re}P_z$ and $\mathrm{Re}P_y$ in Fig. 3 show that the counter-directed Poynting vectors in the guiding plasmonic layer and surrounding it dielectric layers form vortices of power flow of SPPs. Vorticity of the SPP power flow at an air interface of *lossless* plasma half-space had been mentioned first in [30] but its effect on the abnormally high losses of SPPs was not recognised then.

The properties of SPPs and their Poynting vector distributions are strongly influenced by thicknesses of dielectric layers, $a_1$ and $a_2$. At $k_0 a_{1,2} \gg 1$, SPPs are forward waves, and their power flows in the dielectric layers is greater than in the plasmonic layer, as evident in Fig. 3. When the dielectric layer at the guiding interface becomes thinner, SPP in the *lossless* plasmonic layer turns into a backward wave SPP'$_3$ shown in Fig. 2. However, losses qualitatively alter the eigenwave properties in the proximity of SPP resonance. At the result, the backward wave SPP'$_3$ turns into a complex wave SPP$_3$, which is very strongly attenuated at $\omega > \omega_{r2}$ due to its high losses in the plasmonic layer. This effect is illustrated by Fig. 2, where the attenuation constant of backward SPP$_3$ exceeds the propagation constant at $\omega > \omega_{r2}$.

As frequency $\omega$ approaches the SPP resonances at $\omega_{r1}$ and $\omega_{r2}$, the SPPs slow down. Their $\mathrm{Re}\,\gamma$ and $\mathrm{Im}\,\gamma$ grow and vorticity of the power flow increases. At the cut-off frequencies, the oppositely directed power flows become equal in the plasmonic and dielectric layers, and vortices of Poynting vectors are trapped. Thus, vorticity the Poynting vector at the layer interfaces causes the anomalous losses of SPPs at their resonance cut-off.

## IV. MAGNETOPLASMONS IN MAGNETISED PLASMA LAYER

When a biasing dc magnetic field $H_0$ is applied along the *x*-axis in the planar structure shown in Fig. 1, relative permittivity of a plasma layer is described by the tensor

$$\boldsymbol{\varepsilon}_g(\omega) = \mathbf{x}\cdot\mathbf{x}\varepsilon_p(\omega) + (\mathbf{I}-\mathbf{x}\cdot\mathbf{x})\varepsilon_t(\omega) + j(\mathbf{x}\times\mathbf{I})\varepsilon_a(\omega) \tag{7}$$

where
$$\varepsilon_p(\omega) = \varepsilon_L\left(1-\frac{\omega_p^2}{\omega\Omega}\right),\ \varepsilon_t(\omega) = \varepsilon_L\left(1-\frac{\omega_p^2\Omega}{\omega\left[\Omega^2-\omega_c^2\right]}\right),\ \varepsilon_a(\omega) = \varepsilon_L\frac{\omega_p^2\omega_c}{\omega\left[\Omega^2-\omega_c^2\right]}, \tag{8}$$

$\varepsilon_L$ is the background permittivity, $\omega$ is angular frequency, $\Omega = \omega - j\nu$, $\nu$ and $\omega_p$ are the collision and plasma frequencies, $\omega_c = \kappa H_0$ is cyclotron frequency and $\kappa \approx 17.588$ MHz/Oe.

The lowest eigen modes in a magnetised plasma layer are TE and TM waves. TE waves are the ordinary waves, which are not affected by gyrotropy of the plasma layer. Therefore, only TM waves, which exhibit the resonance and nonreciprocal behaviour, are considered here. The DE of the TM waves with the propagator $\exp\{j(\omega t - \gamma z)\}$ can be represented in the form similar to that for SPPs

$$L_1(\gamma)L_2(\gamma) + \left[\frac{\beta_e}{\sinh(\beta_e a_0)}\right]^2 = 0 \tag{9}$$

where $L_m(\gamma) = \varepsilon_e V_m + \beta_e \coth(\beta_e a_0) - (-1)^m \gamma \frac{\varepsilon_a}{\varepsilon_t}$, $m=1,2$; $V_m$ are defined in (2); $\beta_e = \sqrt{\gamma^2 - k_0^2 \varepsilon_e(\omega)}$,

$\varepsilon_e(\omega) = \varepsilon_t(\omega) - \frac{\varepsilon_a^2(\omega)}{\varepsilon_t(\omega)} = \varepsilon_L\left(1-\frac{\omega_p^2(\omega\Omega-\omega_p^2)}{\omega\left[\Omega(\omega\Omega-\omega_p^2)-\omega\omega_c^2\right]}\right)$ is the effective permittivity, $k_0$ and $\gamma$ are free space and longitudinal wavenumbers. It is noteworthy that in the absence of magnetic bias $\omega_c = 0$ and $\varepsilon_a = 0$. This results in $L_m(\gamma)$ being reduced to $K_m(\gamma)$ and DE (9) becoming (2) for SPPs in isotropic plasmonic layer.

Spectrum of TM waves in a magnetised plasma layer includes the dynamic waves and MPs [9]-[12]. The dynamic waves are the conventional surface waves guided by the central layer at $\mathrm{Re}\varepsilon_e(\omega) > \max(\varepsilon_1, \varepsilon_2)$ only. This condition is satisfied at frequencies $\omega_{1u} < \omega < \omega_{qu}$ and $\omega > \omega_{2u} > \omega_{qu}$, where $\omega_{1u} = \sqrt{\omega_p^2 + (\omega_c/2)^2} - \omega_c/2 < \omega_p$, $\omega_{qu} = \sqrt{\omega_p^2 + \omega_c^2}$ and $\omega_{2u} = \sqrt{\omega_p^2 + (\omega_c/2)^2} + \omega_c/2$. At $\omega < \omega_{1u}$ and $\omega_{qu} < \omega < \omega_{2u}$, $\mathrm{Re}\varepsilon_e(\omega)$ is negative, and only MPs are guided by the magnetised plasma layer.

Nonreciprocity is the distinctive feature of MPs. In DE (9) it is described by the last term of $L_m(\gamma)$ dependent on $\gamma\varepsilon_a$. In the case of identical dielectric layers, $V_1 = V_2$, wavenumbers of the oppositely directed waves are the same. But the field distributions differ due to the nonreciprocal field displacement dependent on sign$\gamma$. MPs exist in the two frequency bands and have the resonance cut-offs at the upper bound of each band. In these two bands, MPs travelling in the same direction are guided by the opposite interfaces of the magnetised plasma layer due to the opposite signs of $\mathrm{Re}\{\varepsilon_a(\omega)/\varepsilon_t(\omega)\}$ [12]. For example, if a MP is attached to the upper interface of the plasma layer shown in Fig. 1 at $\omega < \omega_{1u}$, MP of the same direction is displaced to the lower interface in the frequency band $\omega_{qu} < \omega < \omega_{2u}$.

As frequency $\omega$ approaches the MP resonances, $\omega_{1u}$ or $\omega_{2u}$, the MP wavenumbers grow similar to those of SPPs. At $|\gamma| > k_0\sqrt{\max(\varepsilon_{1,2}, \varepsilon_L)}$, the asymptotic solution of DE (9) has the same form as (4), where $\varepsilon_p$ must be replaced by $\varepsilon_m^c = \left[\varepsilon_t + (-1)^m \varepsilon_a\right]$. Then the propagation and attenuation constants of MPs are approximated as follows

$$\begin{aligned}\mathrm{Re}\,\gamma_m &= k_0\sqrt{\frac{\varepsilon_m}{2}\left|\frac{\varepsilon_m^c}{\varepsilon_m^c+\varepsilon_m}\right|(1+\kappa_m)} + O\left(\frac{k_0^2}{|\gamma_m|^2}\right) \\ \mathrm{Im}\,\gamma_m &= |\mathrm{Re}\,\gamma_m|^3 \frac{2\,\mathrm{Im}(\varepsilon_m^c)}{\left[k_0|\varepsilon_m^c|(1+\kappa_m)\right]^2} + O\left(\frac{k_0^2}{|\gamma_m|^2}\right)\end{aligned} \tag{10}$$

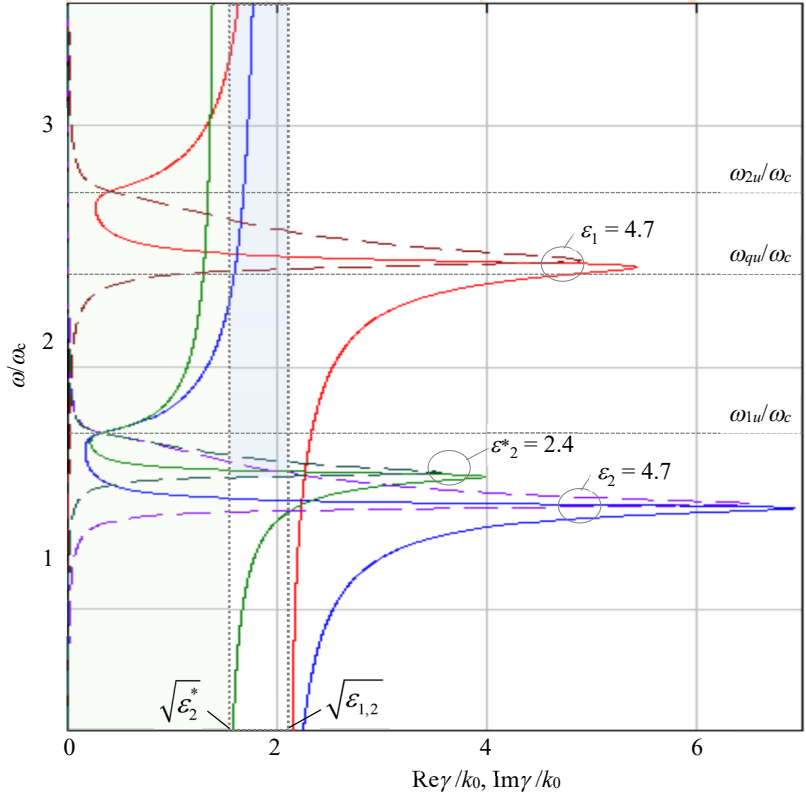

Fig. 4. Asymptotic dispersion characteristics of magnetoplasmons in the three-layer structure of Fig. 1. Propagation constants (solid lines) and attenuation constants (dashed lines) in the cases of dielectric layers with the same permittivities, $\varepsilon_1 = \varepsilon_2 = 4.7$ (red and blue lines) and different permittivities with $\varepsilon_1 = 4.7$, $\varepsilon^*_2 = 2.4$ (red and green lines). In the shaded areas to the left from dotted lines, where $\mathrm{Re}\,\gamma/k_0 < \sqrt{\varepsilon^*_2}$ or $\sqrt{\varepsilon_{1,2}}$ no interface waves are guided by the plasmonic layer. Parameters of the plasmonic layer: $\varepsilon_L = 13.1$, $\omega_p = 2.17\omega_c$, $\nu = 0.02\omega_c$.

where $\varepsilon^c_m$ depends on frequency $\omega$ and $\kappa_m = \dfrac{|\varepsilon^c_m|^2 + \varepsilon_m \mathrm{Re}\,\varepsilon^c_m}{|\varepsilon^c_m||\varepsilon^c_m + \varepsilon_m|}$, $m = 1, 2$. It is important to note that in the proximity of the MP resonances $\kappa_m = \dfrac{\nu}{\omega + (-1)^m \omega_c}\left(1 + \dfrac{\varepsilon_L}{\varepsilon_m}\right)\left[1 + O\left(\dfrac{\nu^2}{\omega^2}\right)\right]$ and $\kappa_m$ are small at $\nu \ll \omega \pm \omega_c$, similar to SPPs.

The dispersion characteristics of MPs in the proximities of their resonances are shown in Fig. 4. They illustrate two cases permittivities of dielectric layers are (a) the same, $\varepsilon_1 = \varepsilon_2 = 4.7$, and (b) different, $\varepsilon_1 = 4.7$, $\varepsilon^*_2 = 2.4$. The propagation and attenuation of MPs in Fig. 4 demonstrate that in the proximities of the resonance cut-offs, attenuation constants $\mathrm{Im}\,\gamma_m$ grow much faster than the propagation constants $\mathrm{Re}\,\gamma_m$ as predicted by (10). It is noteworthy that permittivity of a dielectric layer at the guiding interface influences the cut-off frequencies of MPs, and they slightly increase at lower $\varepsilon_{1,2}$. At frequencies above the MP resonances, the dispersion curves of TM waves are in the shaded areas at the left from the dotted lines where only the fast waveguide modes of the dielectric layers exist. These waves are guided by the dielectric layers and are not the bound to the plasma layer.

The fact that MP losses in (10) grow at the same rate as SPPs in (4), i.e., $\mathrm{Im}\,\gamma_m \sim (\mathrm{Re}\,\gamma_m)^3$, suggests that dissipation of MPs is also the result of their power flow vorticity. The Poynting vector distribution of MPs in (11) shows that it differs from that for SPPs by the dependence on $\varepsilon_a/\varepsilon_t$

$$P_z(y) = \frac{\eta_0}{k_0} Q^2 \, \text{Re} \begin{cases} \frac{\gamma}{\varepsilon_m} \left[ \left(1-(-1)^m W_e\left(\varepsilon_e, \frac{\varepsilon_a}{\varepsilon_t}\right)\right) \frac{\cosh\beta_m\left(a_m + a_c + (-1)^m y\right)}{\cosh\beta_m a_m} \right]^2, & y(-1)^{m-1} > a_c, \ m=1,2 \\ \frac{\gamma}{\varepsilon_e} \left[1 + \frac{\varepsilon_a}{\gamma\varepsilon_t} S\left(y, W_e\left(\varepsilon_e, \frac{\varepsilon_a}{\varepsilon_t}\right)\right)\right] \left[\frac{\cosh\beta_p y}{\cosh\beta_p a_c} + W_e\left(\varepsilon_e, \frac{\varepsilon_a}{\varepsilon_t}\right) \frac{\sinh\beta_p y}{\sinh\beta_p a_c}\right]^2, & |y_c| \leq a_c \end{cases} \quad (11)$$

where $\eta_0$, $k_0$, $Q$, $\varepsilon_e(\omega)$ are defined in (5), (9), $S(y,W) = \beta_p \dfrac{W + \tanh(\beta_p a_c)\tanh(\beta_p y)}{\tanh(\beta_p a_c) + W\tanh(\beta_p y)}$, and nonreciprocity of the MP power flow is described by the terms dependent on $W_e(\varepsilon_e, \varepsilon_a/\varepsilon_t)$ defined in (6).

Poynting vector distribution $P_z(y)$ in (11) shows that at $\text{Re}\varepsilon_e(\omega) < 0$, power flows of MPs inside and outside plasma layer are counter-directional, similar to SPPs in the isotropic plasma layer. Therefore, vorticity of the power flow is the main propagation mechanisms of the MPs responsible for their anomalous losses. It is necessary to note that despite the similarities in the power flows of SPPs and MPs, Poynting vector distribution is more intricate due to gyrotropy of the magnetised plasma layer [31], [32].

Nonreciprocity is the distinctive property of MPs. It manifests itself not only in asymmetry of the field and power flow distributions but also in the nonreciprocity of the cut-off frequencies. Therefore, the propagation and attenuation constants of MPs, displaced to the opposite interfaces of the magnetised plasma layer, differ when the dielectric layers are not the same. It is necessary to note that vorticity of the power flow is stronger in MPs than in SPPs due to the effect of the nonreciprocal field displacement. As the result, losses of MPs are higher than SPPs, despite the *cubic* relation between their attenuation and propagation constants remains the same.

## V. Spin Waves in Ferrimagnetic Layers

Let us consider a ferrimagnetic layer located in the middle of a planar structure shown in Fig. 1. It is magnetised to saturation along the *x*-axis (Voigt configuration) and characterised by a scalar relative permittivity $\varepsilon_g$ and Polder tensor of relative permeability $\boldsymbol{\mu}_g$ [33]

$$\boldsymbol{\mu}_g = \mathbf{x} \cdot \mathbf{x} + \mu_t (\mathbf{I} - \mathbf{x} \cdot \mathbf{x}) - j\mu_a (\mathbf{x} \times \mathbf{I}) \quad (12)$$

where $\mu_t = \dfrac{\Omega_\perp^2 - \omega^2}{\Omega_H^2 - \omega^2}$, $\mu_a = \dfrac{\omega\omega_M}{\Omega_H^2 - \omega^2}$, $\Omega_H = \eta(H_0 - j\Delta H)$, $\omega_M = \eta 4\pi M_s$, $\Omega_B = \Omega_H + \omega_M$, $\Omega_\perp = \sqrt{\Omega_H \Omega_B}$, $H_0$ is internal DC magnetic bias, $\Delta H$ is the ferrimagnetic resonance linewidth, $4\pi M_s$ is the saturation magnetization, and the gyromagnetic ratio $\eta = 2.8$ MHz/Oe.

Spectrum of eigenwaves in the ferrimagnetic layer with the tensor permeability $\boldsymbol{\mu}_g$ includes TE and TM waves. TM waves with the field components $H_x$, $E_y$, $E_z$ are the ordinary waves, which are not affected by the layer gyrotropy. Therefore, they are not considered here. TE waves with the field components $E_x$, $H_y$, $H_z$ are the extraordinary waves, which strongly interact with the ferrimagnetic medium described by the tensor permeability $\boldsymbol{\mu}_g$. The spectrum of TE waves includes the dynamic waves and surface spin waves (SSWs). Properties of the TE waves have been extensively explored in the literature and are used in the nonreciprocal devices [14]-[17], [34], [35]. This is particularly concerned of the dynamic waves, which are the workhorse of the passive ferrite devices such as circulators, isolators and phase shifters, etc., see e.g., [36]-[40]. Applications of spin waves has been limited by their high losses. Despite significant efforts in mitigating SW losses, they remain high. It is shown below that the attenuation constants of SSWs are proportional to a cube of the propagation constant due to vorticity of the power flow, similar to SPPs and MPs.

Let us consider the TE waves with the propagator $\exp\{j(\omega t - \gamma z)\}$. Their DE can be presented in the form similar to the DE for SPPs and MPs

$$M_1(\gamma) M_2(\gamma) - \left[\frac{\beta_g}{\sinh(\beta_g a_0)}\right]^2 = 0 \quad (13)$$

where $M_m(\gamma) = \mu_e U_m + \beta_g \coth(\beta_g a_0) + (-1)^m \gamma \dfrac{\mu_a}{\mu_t}$, $U_m = \beta_m \coth(\beta_m a_m)$, $\beta_m = \sqrt{\gamma^2 - k_0^2 \varepsilon_m}$, $m=1,2$; $\beta_g = \sqrt{\gamma^2 - k_0^2 \varepsilon_g \mu_e}$, $\mu_e = \mu_t - \mu_a^2/\mu_t$; $k_0$ and $\gamma$ are the free space and the longitudinal wavenumbers.

The solutions of DE (13) include the dynamic waves and SSWs. The dynamic waves are the conventional surface waves, which exist only at $\mu_t \cdot \mu_e > 0$, i.e., at frequencies $\omega < \omega_H$ and $\omega > \omega_B$, where $\omega_H = \text{Re}\Omega_H$ and $\omega_B = \mathrm{Re}\Omega_B$ are the ferrimagnetic resonance and plasma frequencies [33]. In contrast to the dynamic waves, SSWs are the IWs. They exist in the finite frequency band $\omega_H < \omega < \omega_B$ only, and at $a_{1,2} > 0$ experience the high frequency cut-off at the SSW resonance $\omega_{sw} = (\omega_H + \omega_B)/2$. Similar to MPs, SSWs are the nonreciprocal waves. Their nonreciprocity is described by the last term in $M_m(\gamma)$ proportional to $\gamma\mu_a/\mu_t$. In the case of identical dielectric layers, $U_1 = U_2$ and wavenumbers of the oppositely directed SSWs are the same. But their field distributions differ due to nonreciprocity of their field displacement, which depends on the propagation direction, i.e., the sign$\gamma$. Examples of the dispersion characteristics and field patterns of SSWs in the planar structure of Fig. 1 are shown in Fig. 5 at several thicknesses of the dielectric layers.

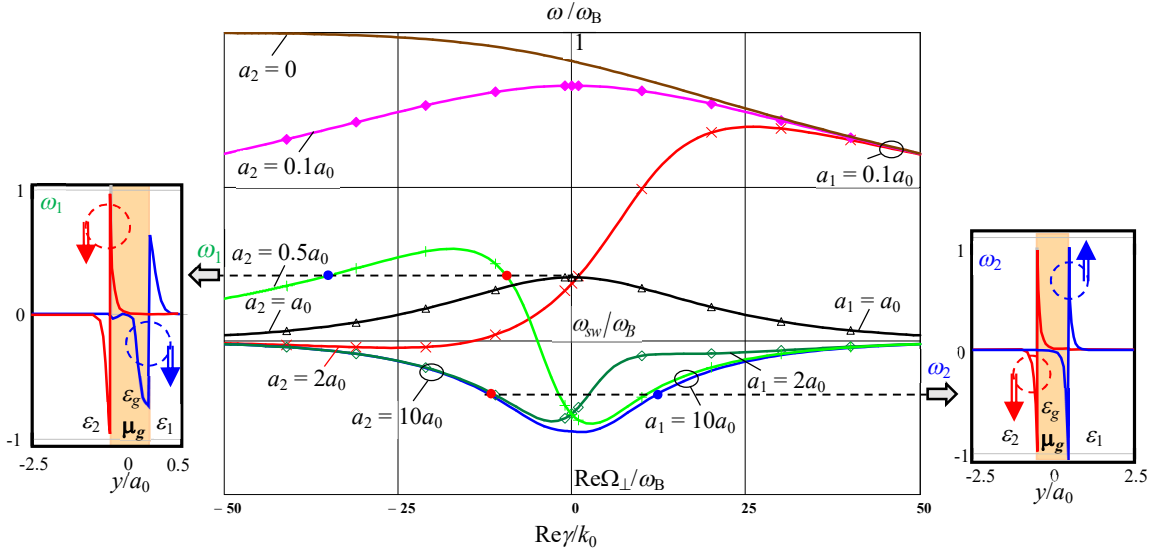

Fig. 5. Dispersion characteristics of spin waves (SWs) at different thicknesses of dielectric layers with the permittivities $\varepsilon_1 = \varepsilon_2 = 1$. Cross-sectional distributions of Poynting vector of SWs are shown in the side panels at frequency $\omega_2$ for forward waves ($a_1 = a_2 = 10a_0$) and at frequency $\omega_1$ for a combination of forward and backward waves ($a_1 = 10a_0$, $a_2 = 0.5a_0$). Red dots on the dispersion diagram and red lines in the field patterns are for the waves travelling in the positive direction of the z-axis, and the blue dots and lines are for the waves travelling in the negative direction of the z-axis. Arrows show the directions of the phase velocities.

The properties of SSWs in the presence of magnetic losses were studied in detail in [15]-[17], and it was found that at frequencies $\omega$ close to $\omega_{sw}$, the wavenumbers of the SSWs grow similar to those of SPPs and MPs. The asymptotic solutions of DE (13) at $|\gamma| \gg k_0\sqrt{\varepsilon_g}$ can be also represented in the form of (10) where $\varepsilon_m^c$ and $\kappa_m$ are replaced by $\mu_{cr} = \mu_t + \mu_a$ and $\xi = \dfrac{|\mu_{cr}|^2 + \mathrm{Re}\,\mu_{cr}}{|\mu_{cr}||\mu_{cr}+1|}$, respectively. Then the asymptotic expansions of the SSW propagation and attenuation constants are approximated as follows

$$\mathrm{Re}\,\gamma = k_0\sqrt{\left|\frac{\mu_{cr}}{\mu_{cr}+1}\right|(1+\xi)} + O\left(\frac{k_0^2\varepsilon_g}{|\gamma|^2}\right)$$

$$\mathrm{Im}\,\gamma = |\mathrm{Re}\,\gamma|^3 \frac{\mathrm{Im}(\mu_{cr})}{2\left[k_0|\mu_{cr}|(1+\xi)\right]^2} + O\left(\frac{k_0^2\varepsilon_g}{|\gamma|^2}\right) \qquad m = 1, 2 \qquad (14)$$

where $\mu_{cr}$ and $\xi$ are frequency dependent. Then at the SSW resonance frequency, $\xi = \dfrac{4\delta\omega_H}{\omega_M}\left[1 + O\left(\dfrac{\delta\omega_H^2}{\omega_M^2}\right)\right]$ and $\xi \ll 1$ at $\delta\omega_H \ll \omega_M$. The asymptotic expansions of $\gamma$ in (14) shows that $\mathrm{Re}\,\gamma$ and $\mathrm{Im}\,\gamma$ do not depend on $\mathrm{sign}(\mathrm{Re}\,\gamma)$ in the proximity of the SSW resonance. This means that SSW nonreciprocity is small at $\mathrm{Re}\,\gamma \gg k_0\sqrt{\varepsilon_g}$ because the frequencies of SW resonances are weakly affected the permittivities of dielectric layers, $\varepsilon_{1,2}$. This is illustrated by the dispersion curves of SSWs in Fig. 5 at $a_{1,2} \geq 2a_0$. Therefore, despite the nonreciprocity, the propagation and attenuation constants of SSWs remain practically the same for both propagation directions at $\mathrm{Re}\,\gamma \gg k_0\sqrt{\varepsilon_g}$.

An important feature of the asymptotic dispersion relations in (14) is that the SSW attenuation constants grow at the same rate, $\text{Im}\gamma \sim (\text{Re}\gamma)^3$, as that for SPPs and MPs, see (4) and (10). This means that the mechanisms of the SSW propagation and losses should be similar to the other types of the IWs and be related to the vortices of power flow at the layer interfaces. Indeed, distribution of the longitudinal component of the normalised Poynting vector in dielectric and ferrimagnetic layers shows that it has the same form as that for the magnetised plasma layer considered earlier, viz.

$$P_z(y) = Q^2 \frac{k_0}{\eta_0} \begin{cases} \text{Re}\left\{\gamma\left[\left(1-(-1)^m W_m\left(\mu_e, \frac{\mu_a}{\mu_t}\right)\right)\frac{\sinh \beta_m \left(a_m + a_c + (-1)^m y\right)}{\sinh \beta_m a_m}\right]^2\right\}, & y(-1)^{m-1} > a_c, \; m=1,2 \\ \text{Re}\left\{\frac{\gamma}{\mu_e}\left[1 - \frac{\mu_a}{\gamma \mu_t} S\left(y, W_m\left(\mu_e, \frac{\mu_a}{\mu_t}\right)\right)\right]\left(\frac{\cosh \beta_p y}{\cosh \beta_p a_c} + W_m\left(\mu_e, \frac{\mu_a}{\mu_t}\right)\frac{\sinh \beta_p y}{\sinh \beta_p a_c}\right)^2\right\}, & |y| \le a_c \end{cases} \quad (15)$$

where $\eta_0$, $k_0$, $\mu_e$, $\mu_t$, $\mu_a$ and $S(y, W)$ are defined in (5), (12) and (11), and nonreciprocity of the SW power flow is determined by $W_m(\mu_e, \mu_a/\mu_t)$ which depends on $\gamma$

$$W_m(\upsilon, \varpi) = \frac{\upsilon\left[U_2(\gamma) - U_1(\gamma)\right] + 2\gamma\varpi}{\upsilon\left[U_2(\gamma) + U_1(\gamma)\right] + 2\beta_p \coth(\beta_p a_c)} \quad (16)$$

Comparison of (15) and (16) with (11) shows that similar to MPs, power flows of SSWs in the ferrimagnetic layer and surrounding it dielectric layers are directed oppositely. This is illustrated by power flow distributions in the side panels in Fig. 5. Indeed, they show that vorticity of the power flow at the guiding interfaces is responsible for the high losses of SSWs [17]. Similar to MPs, SSW power flows are asymmetrically shifted to the guiding interfaces due to the nonreciprocal field displacement in the ferrimagnetic layer even when the surrounding dielectric layers are identical. Right panel in Fig. 5 illustrates the power flow distribution at frequency $\omega_2$ in the symmetric structure with $a_1 = a_2 = 10a_0$. In this case, SSWs are attached to the guiding interfaces, and power flows in the dielectric layers are marginally larger than in the ferrimagnetic layer. This results in the normal dispersion of SSWs. When thicknesses of dielectric are layers reduced, e.g., $a_1 = 10a_0$ and $a_2 = 0.5a_0$ in the left panel of Fig. 5, power flows of the waves travelling in opposite directions of the $z$-axis are still displaced to the opposite interfaces at frequency $\omega_1$. However, at the right interface, power flow of the SSW inside the ferrimagnetic layer becomes larger than in the adjacent dielectric layer. This results in the anomalous dispersion (blue dot on the dispersion diagram) and the higher losses of the backward SW.

At other directions of the magnetic bias $\mathbf{H_0}$, the field distributions of the SWs change. Namely, the SW travelling along the direction of the in-plane magnetisation ($\mathbf{H_0}\|0z$) are volume backward waves, and at the magnetic bias normal to the layer surface ($\mathbf{H_0}\|0y$), the SWs are forward volume waves. But their main features, cubic relation between their attenuation and propagation constants and vortices of power flow remain unchanged. These common properties of the SWs are the result of their resonance cut-off at the intrinsic ferrimagnetic resonances of the central layer. Despite the differences in the types of SW dispersion and field distributions (volume or interface) inside the ferrimagnetic layer, the mechanisms of their cut-off remain similar for both SSWs and volume SWs as the SWs of any type experience the resonance absorption when vortices of their power flow are trapped.

## VI. Conclusion

The dispersion and attenuation properties of IWs and their relations to the power flow distributions have been discussed with the examples of SPPs, MPs and SSWs. It is shown that these IWs are the waves of hyperbolic type, and the fundamental mechanisms of their propagation and dissipation are similar. This is evidenced by that similarity of the dispersion equations (2), (9), (13) for all three types of the IWs. The main results of this work are
- the attenuation of the IWs grows at the rate of $[\text{Re}\gamma(\omega)]^3$ in the proximity of the resonance cut-offs, and
- the high losses of the IWs are caused by the vorticity of Poynting vector inherent to hyperbolic media.

The mechanisms of losses, discussed in the paper by the examples of SPPs, MPs and SSWs, are common for any slow waves in the hyperbolic media with intrinsic resonances at finite frequencies. It is necessary to stress that the nonreciprocity of MPs and SWs is asymptotically small in the proximity of their resonance cut-offs and does not affect the cubic relation between the propagation and attenuation constants in eqns. (10) and (14).